\documentclass{achemso}

\SectionNumbersOn

\usepackage{graphicx}
\usepackage{color}
\usepackage{threeparttable}
\usepackage{url}
\usepackage{amsmath}

\allowdisplaybreaks     

\makeatletter

\newcommand{\Runum}[1]{\expandafter\@slowromancap\romannumeral #1@}
\makeatother

\title{Libcint: An efficient general integral library for Gaussian basis
functions}
\author{Qiming Sun}
\affiliation{Department of Chemistry, Princeton University, Princeton NJ 08544}
\email{osirpt.sun@gmail.com}

\begin{document}

\maketitle

\begin{abstract}
An efficient integral library Libcint was designed to automatically
implement general integrals for Gaussian-type scalar and spinor basis
functions.
The library can handle arbitrary integral expressions on top of $\mathbf{p}$,
$\mathbf{r}$ and $\sigma$ operators with one-electron overlap and nuclear
attraction, two-electron Coulomb and Gaunt operators.
Using a symbolic algebra tool, new integrals are derived and translated to C
code programmatically.
The generated integrals can be used in various types of molecular properties.
In the present work, we computed the analytical gradients and NMR shielding
constants at both non-relativistic and four-component relativistic
Hartree-Fock level to demonstrate the capability of the integral library.
Due to the use of kinetically balanced basis and gauge including atomic
orbitals, the relativistic analytical gradients and shielding constants
requires the integral library to handle the fifth-order electron repulsion
integral derivatives.
The generality of the integral library is achieved without losing efficiency.
On the modern multi-CPU platform,
Libcint can easily reach the overall throughput being many times of the I/O
bandwidth.  On a 20-core node, we are able to achieve an average output 7.9
GB/s for C$_{60}$ molecule with cc-pVTZ basis.
\end{abstract}

\section{Introduction}
In computational chemistry,
evaluation of integrals is the ground of modeling the molecular electronic
structure and various types of molecular properties.
Because of the importance of integrals, 
a considerable number of researches have been devoted on the efficient algorithm
to evaluate the two-electron repulsion integrals (ERI) over
Gaussian functions\cite{Boys1950,MichelDupuis1976,Rys1983,Pople1978,McMurchie1978,
Obara1986,Obara1988,Schlegel1982,Klopper1992,Pople1988,Pople1989,Pople1990,
Pople1991,Lindh1991,Lindh1993,Dupuis2001,Ishida1991,Ishida1996}
since Boys' work\cite{Boys1950} in 1950.
Grounded on the Rys-quadrature method which was developed by Dupuis, Rys, and
King (DRK)\cite{MichelDupuis1976,Rys1983},
Pople and Hehre\cite{Pople1978} developed an efficient method for highly contracted $s$
and $p$ functions.
To reduce the number of floating-point operations (FLOPS) required by the
quadrature integration technique,
Obara and Saika proposed a recurrence
relation\cite{Obara1986,Obara1988,Schlegel1982} (RR) for the 6D-integral
rather than the 2D-integral of the original DRK's algorithm.
Based on OS's formula,
an early contraction scheme was proposed by
Gill, Head-Gordon and Pople\cite{Pople1988,Pople1989,Pople1990,Pople1991}
(HGP).  Their formula transfered the RR out of the contraction loops
and further reduced FLOPS.
Although OS's method as well as HGP's early contraction scheme required less
FLOPS counts, they are less efficient than DRK's 2D-integral algorithm in some
scenario for low degree of contraction due to the complex formula in their
algorithm\cite{Lindh1991}.
Lindh, Ryu, and Liu\cite{Lindh1991,Lindh1993} first noticed this problem and
proposed a compromised scheme.
Dupuis and Marquez\cite{Dupuis2001} combined the advantages of DRK, HGP etc.
algorithms to optimize the FLOPS counts.
Besides these applications,
Ishida's ACE algorithm\cite{Ishida1991,Ishida1996} also achieved attractive
FLOPS counts.

Most of the existed integral algorithms focused on the FLOPS counts which only
provide the theoretical computing efficiency.
It remains a challenge to implement new types of integrals cost-effectively in
both the human labour and the real computational efforts.
One source of the new integrals is the calculation of molecular properties
such as \cite{Helgaker2012} response theory, which requires the differentiated
integrals\cite{Visscher2010}.
Perturbation-dependent basis technique\cite{London1937,Ditchfield1972,Schlegel1994},
e.g. gauge including atomic orbitals\cite{London1937,Ditchfield1972}
(GIAO) also introduces the complications on the evaluation of integrals.
Inclusion of relativistic effect is another source that brings new
integrals\cite{Pyykko1988,Helgaker2012,Stanton1984,Cheng2014,LanCheng2009a}.
It is very common in relativistic quantum chemistry to evaluate high-order
integrals, e.g.
Breit-Pauli Hamiltonian\cite{Breit1929,Kutzelnigg2000} introduces many
one-electron and two-electron operators in terms of the product of
$\mathbf{p}$ and $\mathbf{r}$ operators.
One basic assumption in four-component relativistic theory is the balanced
treatment of large and small
components\cite{Stanton1984,Dyall1990,Ishikawa1983,YunlongXiao2007}.
It results in various types of kinetic (magnetic) balance
conditions\cite{Stanton1984,Yanai2002,Kelley2013,Cheng2014,QimingSun2011,LanCheng2009a}
and the corresponding one-electron and two-electron $j$-adapted (spinor)
integrals.
It is a heavy task to manually implement new code to efficiently evaluate every new
type of integrals.

To implement an efficient integral program, the architecture of modern
computer is another important subject that should be taken into
account\cite{AlexeyTitov2012,Gordon2012a}.
The efficiency of an algorithm is not merely determined by the necessary
FLOPS.
A well optimized code can take full advantage of computer architecture, such
as the single instruction multiple data (SIMD) units, to achieve instruction
level parallelization\cite{Blelloch1994}.
Data locality can also affects program efficiency since
the modern computer hardware favours simple data structure which are local and
aligned in the memory. 
As such, it is non-trivial to translate an integral algorithm to a real-world
efficient implementation\cite{Yasuda2007,Gordon2010,Luehr2011,AlexeyTitov2012,Gordon2012a}.

Targeting to efficiently provide new integrals, an open-source and general
purposed integral library Libcint\cite{Libcint2014} was designed.
In this library, a built-in symbolic algebra system can parse the integral
expression which is the polynomial of $\mathbf{p}$ operator, $\mathbf{r}$
operator and Pauli matrices $\sigma$ and decompose the expressions to the
basic Cartesian integrals.
The basic Cartesian ERI are evaluated with DRK's algorithm.  There are two
reasons we made this choice.  ({\romannumeral 1}) the intermediates for the
derived integrals and the basic ERIs have the similar structure in the DRK's
2D-integral framework.  It allows us to reuse most of the code which has been
highly optimized for the basic ERIs.  It also reduces the complexity
of the code generator.
({\romannumeral 2}) we observed that the data structure of DRK's algorithm
shows high locality which is easy to be fit into the CPU Cache structure.
To cope with $j$-adapted spinor integrals for the four-component and
two-component relativistic theory,
the symbolic program can pick a proper function to assemble the intermediate
Cartesian integrals.

In this paper, we describe the detail of the integral generation algorithm in
Section \ref{algorithm}.
As a numerical example, we computed analytical nuclear gradients and NMR
shielding constants for Cr(CO)$_6$ and UF$_6$ molecule.  They are presented in
Section \ref{numexamples}.
The performances of spherical and spinor integrals for ethane molecule with
double, triple and quadruple zeta bases are tested and compared in Section
\ref{performance}.

\section{Algorithm}
\label{algorithm}
The evaluation of integral can be divided into two separate steps.
First is to calculate all kinds of primitive intermediate Cartesian
integrals.  Second step is to assemble and contract the
intermediates, then transform them to the real spherical or spinor
representations.
In the following paragraphs, we will use the superscripts $C$, $S$ and $J$ to
denote the integrals in the Cartesian, spherical and $j$-adapted spinor
representations.

In the first step, we implemented a symbolic algebra program to parse the
integral expression and formulate the intermediates.
As shown in Table \ref{tab:ops}, the supported operators are classified into
three classes: scalar, vector, and compound.
The scalar and vector operators are the basic operators.
The compound operators can be expressed in terms of the basic operators.
In order to handle the Pauli matrices, we used quaternion
\begin{gather*}
  \sigma_x =
  \begin{pmatrix}
    0 & 1 \\
    1 & 0
  \end{pmatrix}, \quad
  \sigma_y =
  \begin{pmatrix}
    0 &-i \\
    i & 0
  \end{pmatrix}, \quad
  \sigma_z =
  \begin{pmatrix}
    1 & 0 \\
    0 &-1
  \end{pmatrix}, \quad
  \mathbf{1}_{2\times2} =
  \begin{pmatrix}
    1 & 0 \\
    0 & 1
  \end{pmatrix}
  \label{eq:quaternion}
\end{gather*}
as the fundamental structure to represent the scalar and vector operators.
A scalar operator can be written as
\begin{equation*}
  q = q^x\sigma_x + q^y\sigma_y + q^z\sigma_z + q^1\mathbf{1}_{2\times2}
\end{equation*}
A vector operator contains three quaternions
\begin{equation*}
  \vec{q} = q_x\mathbf{e}_x + q_y\mathbf{e}_y + q_z\mathbf{e}_z
\end{equation*}
Although many zeros might be introduced due to the quaternion representation,
the evaluation of the quaternion expression is simple.  An valid quaternion
expression can only have three kinds of basic contractions: dot product,
\begin{gather}
  \vec{q}_a \cdot \vec{q}_b
  = q_{a,x}q_{b,x} + q_{a,y}q_{b,y} + q_{a,z}q_{b,z}
  \label{eq:ctr1}
\end{gather}
cross product,
\begin{gather}
  \vec{q}_a \times \vec{q}_b
  = (q_{a,y}q_{b,z}-q_{a,z}q_{b,y})\mathbf{e}_x
  + (q_{a,z}q_{b,x}-q_{a,x}q_{b,z})\mathbf{e}_y
  + (q_{a,y}q_{b,z}-q_{a,z}q_{b,y})\mathbf{e}_z
  \label{eq:ctr2}
\end{gather}
and direct product
\begin{gather}
  q_a \vec{q}_b
  = q_{a}q_{b,x}\mathbf{e}_x
  + q_{a}q_{b,y}\mathbf{e}_y
  + q_{a}q_{b,z}\mathbf{e}_z
  \\
  \vec{q}_a q_b
  = q_{a,x}q_{b}\mathbf{e}_x
  + q_{a,y}q_{b}\mathbf{e}_y
  + q_{a,z}q_{b}\mathbf{e}_z
  \\
\begin{aligned}
  \vec{q}_a \vec{q}_b
  &=q_{a,x}q_{b,x}\mathbf{e}_x \mathbf{e}_x
  + q_{a,x}q_{b,y}\mathbf{e}_x \mathbf{e}_y
  + q_{a,x}q_{b,z}\mathbf{e}_x \mathbf{e}_z \\
  &+q_{a,y}q_{b,x}\mathbf{e}_y \mathbf{e}_x
  + q_{a,y}q_{b,y}\mathbf{e}_y \mathbf{e}_y
  + q_{a,y}q_{b,z}\mathbf{e}_y \mathbf{e}_z \\
  &+q_{a,z}q_{b,x}\mathbf{e}_z \mathbf{e}_x
  + q_{a,z}q_{b,y}\mathbf{e}_z \mathbf{e}_y
  + q_{a,z}q_{b,z}\mathbf{e}_z \mathbf{e}_z
\end{aligned}
  \label{eq:ctr3}
\end{gather}
where the quaternion multiplication $q_a q_b$ can be expanded in terms of the
Dirac relation
\begin{gather}
  \sigma\cdot \mathbf{A}
  \sigma\cdot \mathbf{B}
  = \mathbf{A}\cdot\mathbf{B} + i\sigma\cdot\mathbf{A}\times\mathbf{B}
  \notag\\
  q_a q_b = q_c
  = q_c^x\sigma_x + q_c^y\sigma_y + q_c^z\sigma_z + q_c^1\mathbf{1}_{2\times2}
  \label{eq:qprod}
  \\
  q_c^x = iq_a^y q_b^z - iq_a^z q_a^y + q_a^x q_b^1 + q_a^1 q_b^x \notag\\
  q_c^y = iq_a^z q_b^x - iq_a^x q_a^z + q_a^y q_b^1 + q_a^1 q_b^y \notag\\
  q_c^z = iq_a^x q_b^y - iq_a^y q_a^x + q_a^z q_b^1 + q_a^1 q_b^z \notag\\
  q_c^1 = q_a^x q_b^x + q_a^y q_b^y + q_a^z q_b^z + q_a^1q_b^1    \notag
\end{gather}
However, a special treatment is needed for the Gaunt interactions
\begin{equation*}
  \frac{\alpha_1\cdot\alpha_2}{r_{12}}, \quad
  \alpha =
  \begin{pmatrix}
    0 & \sigma \\
    \sigma & 0
  \end{pmatrix}
\end{equation*}
The two $\alpha$ operators in the Gaunt operator belongs to the different
electrons.  We cannot use Eq. \eqref{eq:qprod} to simplify the dot product.
Instead, it was decomposed to three components.
The three components are calculated separately and summed up at last.

By recursively calling the contractions and quaternion products
\eqref{eq:ctr1} - \eqref{eq:qprod}, we are able to derive the expressions of
all possible Cartesian intermediates for all tensor components.
E.g.  the symbolic program can generate in total six Cartesian intermediates
for $(a~\sigma\times\mathbf{p}b|cd)^J$ which comes with three Cartesian tensor
components
\begin{align*}
  (a~\sigma\times\mathbf{p}b|cd)_x^J: \quad
  -i(a~\nabla_zb|cd)^C\sigma_y, \quad
   i(a~\nabla_yb|cd)^C\sigma_z,
  \label{eq:eg1}\\
  (a~\sigma\times\mathbf{p}b|cd)_y^J: \quad
  -i(a~\nabla_xb|cd)^C\sigma_z, \quad
   i(a~\nabla_zb|cd)^C\sigma_x,
  \\
  (a~\sigma\times\mathbf{p}b|cd)_z^J: \quad
  -i(a~\nabla_yb|cd)^C\sigma_x, \quad
   i(a~\nabla_xb|cd)^C\sigma_y.
\end{align*}

\begin{table}[htp]
  \centering
  \caption{The operators supported by Libcint library}
  \begin{tabular}{llllll}
    \hline
    Operators          & Class & Expression \\
    \hline
    $\nabla_x$         & Scalar   \\
    $\nabla_y$         & Scalar   \\
    $\nabla_z$         & Scalar   \\
    $p_x$              & Compound & $-i\nabla_x$ \\
    $p_y$              & Compound & $-i\nabla_y$ \\
    $p_z$              & Compound & $-i\nabla_z$ \\
    $x$                & Scalar   \\
    $y$                & Scalar   \\
    $z$                & Scalar   \\
    $\sigma_x$         & Scalar   \\
    $\sigma_y$         & Scalar   \\
    $\sigma_z$         & Scalar   \\
    $\frac{1}{|r-R|}$  & Scalar   \\
    $\frac{1}{r_{12}}$ & Scalar   \\
    $\nabla$           & Vector   \\
    $\mathbf{p}$       & Compound & $-i\nabla$ \\
    $\mathbf{r}$       & Vector   \\
    $\sigma$           & Vector   \\
    $\frac{\mathbf{r}-\mathbf{R}}{|r-R|^3}$ & Compound & $-\nabla\frac{1}{|r-R|}$ \\
    $\hat{g}_{\mu\nu}$ & Compound & $i(\mathbf{R}_\mu-\mathbf{R}_\nu)\times \mathbf{r}$ \\
    Gaunt-like & Compound & $\frac{\sigma_1\cdot\sigma_2}{r_{12}}$ & \\
    \hline
  \end{tabular}
  \label{tab:ops}
\end{table}

Next thing the symbolic program did is to translate the expression of
Cartesian intermediates to C code.
Following DRK's method, a Cartesian integral can be evaluated as the
inner product of three two-dimensional integrals $I_x$, $I_y$, $I_z$
with certain weights $w_i$
\begin{equation}
  \mathrm{ERI}
  = \int_0^\infty I_x I_y I_z du = \sum_{i} w_i I_x(i) I_y(i) I_z(i).
  \label{eq:rysum}
\end{equation}
When an integral expression contains $\nabla$ or $\mathbf{r}$ operators, we
need to employ
\begin{equation*}
  \begin{aligned}
    \frac{\partial}{\partial x} \phi_a^x
    &=n_a^x(x-X_A)^{n_a^x-1} e^{-\alpha_a(x-X_A)^2}
    -2\alpha_a(x-X_A)^{n_a^x+1} e^{-\alpha_a(x-X_A)^2}
    \\
    x \phi_a^x
    &=X_A(x-X_A)^{n_a^x} e^{-\alpha_a(x-X_A)^2}
    +(x-X_A)^{n_a^x+1} e^{-\alpha_a(x-X_A)^2}
  \end{aligned}
\end{equation*}
to transfer the four-index 2D integral $I_x^{(ab|cd)}$ to another
four-index 2D integral $\tilde{I}_x^{(ab|cd)}$ which is a linear combination
of a lower and a higher 2D integrals,
\begin{equation}
  \begin{aligned}
    I_x^{(\nabla_x~ab|cd)}
    &= n_a^xI_x^{(a-1~b|cd)} - 2\alpha_a I_x^{(a+1~b|cd)}
    \\
    I_x^{(x~ab|cd)}
    &= X_A I_x^{(ab|cd)} + I_x^{(a+1~b|cd)}
  \end{aligned}
  \label{eq:nablax2d}
\end{equation}
According to these relations, we are able to build the derived 2D integrals
$\tilde{I}_x$in an ``assembling'' subroutine which consumes one $\nabla$ or
$\mathbf{r}$ then form a derived 2D integral once at a time.  If the integral
contains two or more operators, the assembling subroutine needs to be invoked
recursively until all operators are consumed.

It should be noted that the order we applied relations
\eqref{eq:nablax2d} is opposite to the natural order we manipulate the
operators. 
To apply a list of operators to a function,
the natrual order starts from the rightmost operator.
But in the assembling subroutine, relations \eqref{eq:nablax2d} are
invoked from the leftmost operator.
E.g. in terms of the natural order, the leftmost derivative operator of
$\nabla_x x\nabla_x\nabla_x\phi_a^x$ can produce a factor $4\alpha_a^2(n_a^x+3)$
\begin{equation*}
\begin{array}{llllllll}
                     & \nabla_x           & x          & \nabla_x  & \nabla_x   & \phi_a^x \\
\text{momentum index}& n_a^x+2            & n_a^x+3    & n_a^x+2   & n_a^x+1    & n_a^x    \\
\text{factor}        &4\alpha_a^2(n_a^x+3)&4\alpha_a^2 &4\alpha_a^2& -2\alpha_a & 1        \\
\end{array}
\end{equation*}
where the momentum index of the Gaussian function
$\phi_a^x = (x-X_A)^{n_a^x} e^{-\alpha_a(x-X_A)^2}$ stands for the exponential
of the polynomial part $(x-X_A)^{n_a^x}$.
By calling relations \eqref{eq:nablax2d}, left-to-right propagation can
produce the same factor
\begin{equation*}
\begin{array}{llllllll}
& \nabla_x & x & \nabla_x & \nabla_x & I_x^{(ab|cd)} \\
& \multicolumn{5}{l}{I_x^{[1]}(n) = \underline{n_a^x}I_x^{(a-1~b|cd)} - 2\alpha_a I_x^{(a+1~b|cd)}} \\
& & \multicolumn{4}{l}{I_x^{[2]}(n) = I_x^{[1]}(n+1) + X_A I_x^{[1]}(n)
    = \underline{(n_a^x+1)}I_x^{(ab|cd)} + \cdots}\\
& & & \multicolumn{3}{l}{I_x^{[3]}(n) = -2\alpha_a I_x^{[2]}(n+1)
    + n I_x^{[2]}(n-1)
    = \underline{-2\alpha_a(n_a^x+2)}I_x^{(a+1~b|cd)} + \cdots} \\
& & & & \multicolumn{2}{l}{I_x^{[4]}(n) = -2\alpha_a I_x^{[3]}(n+1)
    + n I_x^{[3]}(n-1)
    = \underline{4\alpha_a^2(n_a^x+3)}I_x^{(a+2~b|cd)} + \cdots}
\end{array}
\end{equation*}
Applying similar analysis to all other terms, we found the same observation on
the application orders:  The correct factor can only be produced by the
left-to-right order with the relations \eqref{eq:nablax2d}.

In the second step,
the transformations of Cartesian to spherical or Cartesian to spinor were
hard-coded in the program.
There are eight kinds of transformations for the spinor integrals, which are
arose from the combinations of three conditions:
\begin{itemize}
  \item Which electron to transform.  The four indices in $(ab|cd)^J$ are
    be grouped into two sets $ab$ and $cd$.
  \item Whether the integral expression has Pauli matrices.
    E.g. $(\sigma\mathbf{p}a~\sigma\mathbf{p}b|cd)^J$ contains Pauli matrices,
    but $(a~\sigma\mathbf{p}\sigma\mathbf{p}b|cd)^J$ does not because
    $\sigma\mathbf{p}\sigma\mathbf{p} = p^2$.
  \item Which phase the integral is associated with, 1 or $i$.
    E.g. $(a~\sigma\times\mathbf{p}b|cd)^J$ has a phase factor $i$ from
    operator $\mathbf{p}$.
\end{itemize}
For a given integral expression, the symbolic program needs to identify the
transformation from the above three conditions and choose the proper
transformation subroutines to execute.

\section{Computational examples}
\label{numexamples}
In this section, we present the numerical examples for the integrals
implemented with Libcint library.
We used Pyscf\cite{Libcint2014} program package to call the integral library
and calculate the ground state energy, analytical nuclear gradients and NMR
shielding constants for Cr(CO)$_6$ and UF$_6$ molecule at non-relativistic and
4-component (4C) relativistic (Dirac-Coulomb Hamiltonian) mean field level.
We used cc-pVTZ basis for Cr, C, O and F, Dyall triple-zeta
set\cite{Dyall2002} for U.
In the 4C relativistic calculations, we uncontracted the basis of Cr atom to
get better description of the core electrons.
For the relativistic ground state and nuclear gradients, we employed the
restrict kinetically balanced (RKB) basis sets, which
introduces the $\sigma\cdot\mathbf{p}|a\rangle$ basis functions.
For NMR properties, we employed magnetic-field-dependent basis functions.
They are GIAOs for non-relativistic Hamiltonian
\begin{equation*}
  -\frac{i}{2}\mathbf{B}\times \mathbf{R}_a \cdot\mathbf{r}|a\rangle 
\end{equation*}
and magnetically balanced RMB-GIAOs basis for relativistic Hamiltonian
\begin{align*}
  &-\frac{i}{2}\mathbf{B}\times \mathbf{R}_a \cdot\mathbf{r}|a\rangle \quad 
  \text{for large components} \\
  &(\frac{1}{2}\mathbf{B}\times\mathbf{r}\cdot\sigma
  -\frac{i}{2}\mathbf{B}\times \mathbf{R}_a \cdot\mathbf{r})|a\rangle \quad
  \text{for small components}
\end{align*}
The magnetically balanced basis naturally introduces the dia-magnetic
contributions to the relativistic NMR theory, which is comparable to the
dia-magnetic terms in the non-relativistic calculations.
Since the theory of the relativistic analytic gradients and magnetic
properties is out of the scope of present paper, we refer the readers to the
literatures\cite{Cheng2014,Shiozaki2013,Wang2002a,YunlongXiao2007,Komorovsko2008,Xiao2012}
for more theoretical details.

Table \ref{tab:ints} documents all the 49 types of integrals which are
required in these calculations.
Due to the use of RKB and RMB-GIAO basis, relativistic theory brings more
integrals than that appeared in the non-relativistic theory.
The non-relativistic computation only needs 3 types of two-electron spherical
integrals while the relativistic framework needs 13 types of two-electron
spinor integrals.
Among the 13 types,
$(\nabla \sigma\mathbf{p}a~\sigma\mathbf{p}b|\sigma\mathbf{p}c~\sigma\mathbf{p}d)$
and $(\hat{g}_{ab}\sigma\mathbf{p}a~\sigma\mathbf{p}b|\sigma\mathbf{p}c~\sigma\mathbf{p}d)$
virtually require the fifth order derivative, which causes the relativistic
computation being about 100 times slower than the corresponding
non-relativistic computation (see more discussions in Section \ref{performance}).

\begin{table}
  \centering
  \caption{Integral types for ground state, analytical nuclear gradients
  and NMR shielding constants.}
  \begin{tabular}{lllllll}
    \hline
    Hamiltonian    & Non-relativistic & 4C Dirac-Coulomb \\
    \hline
    ground HF
& $\langle a|b\rangle^S$                                             
        & $\langle a|b\rangle^J$                                                        \\
& $\langle a|\nabla^2 b\rangle^S$                                    
        & $\langle a|\nabla^2 b\rangle^J$                                               \\
& $\langle a|\frac{Z_N}{r_N}|b\rangle^S$                             
        & $\langle a|\frac{Z_N}{r_N}|b\rangle^J$                                        \\
&       & $\langle \sigma\mathbf{p}a|\frac{Z_N}{r_N}|\sigma\mathbf{p}b\rangle^J$        \\
& $(ab|cd)^S$
        & $(ab|cd)^J$                                                                   \\
&       & $(\sigma\mathbf{p}a~\sigma\mathbf{p}b|cd)^J$                                  \\
&       & $(\sigma\mathbf{p}a~\sigma\mathbf{p}b|\sigma\mathbf{p}c~\sigma\mathbf{p}d)^J$ \\
    \hline
    gradients
& $\langle \nabla a|b\rangle^S$                                      
        & $\langle \nabla a|b\rangle^J$                                                        \\
& $\langle \nabla a|\nabla^2 b\rangle^S$                             
        & $\langle \nabla a|\nabla^2 b\rangle^J$                                               \\
& $\langle \nabla a|\frac{Z_N}{r_N}|b\rangle^S$                      
        & $\langle \nabla a|\frac{Z_N}{r_N}|b\rangle^J$                                        \\
&       & $\langle \nabla\sigma\mathbf{p}a|\frac{Z_N}{r_N}|\sigma\mathbf{p}b\rangle^J$         \\
& $\langle a|(\nabla \frac{Z_N}{r_N})|b\rangle^S$                    
        & $\langle a|(\nabla \frac{Z_N}{r_N})|b\rangle^J$                                      \\
&       & $\langle \sigma\mathbf{p}a|\nabla(\frac{Z_N}{r_N})|\sigma\mathbf{p}b\rangle^J$       \\
& $(\nabla a~b|cd)^S$                                                
        & $(\nabla a~b|cd)^J$                                                                  \\
&       & $(\nabla \sigma\mathbf{p}a~\sigma\mathbf{p}b|cd)^J$                                  \\
&       & $(\sigma\mathbf{p}a~\sigma\mathbf{p}b|\nabla c~d)^J$                                 \\
&       & $(\nabla \sigma\mathbf{p}a~\sigma\mathbf{p}b|\sigma\mathbf{p}c~\sigma\mathbf{p}d)^J$ \\
    \hline
    NMR shielding
& $\langle a|\frac{\mathbf{r}\mathbf{r}}{r^3}|b\rangle^S$            
        & $\langle \mathbf{r}\times\sigma a|\frac{\mathbf{r}\times\sigma}{r^3}|b\rangle^J$                            \\
& $\langle a|\frac{\mathbf{r}\times\mathbf{p}}{r^3}|b\rangle^S$      
        & $\langle a|\frac{\mathbf{r}\times\sigma}{r^3}|\sigma\mathbf{p}b\rangle^J$       \\
& $\langle a|\mathbf{r}\times\mathbf{p}|b\rangle^S$                  
        & $\langle \mathbf{r}\times\sigma a|\sigma\mathbf{p}b\rangle^J$                                               \\
&       & $\langle \mathbf{r}\times\sigma a|\frac{Z_N}{r_N}|\sigma\mathbf{p}b\rangle^J$                               \\
& $\langle \hat{g}_{ab}a|b\rangle^S$                 
        & $\langle \hat{g}_{ab}a|b\rangle^J$                                                          \\
& $\langle \hat{g}_{ab}a|\nabla^2 b\rangle^S$        
        & $\langle \hat{g}_{ab}\sigma\mathbf{p}a|\sigma\mathbf{p} b\rangle^J$                         \\
& $\langle \hat{g}_{ab}a|\frac{Z_N}{r_N}|b\rangle^S$ 
        & $\langle \hat{g}_{ab}a|\frac{Z_N}{r_N}|b\rangle^J$                                          \\
&       & $\langle \hat{g}_{ab}\sigma\mathbf{p}a|\frac{Z_N}{r_N}|\sigma\mathbf{p}b\rangle^J$          \\
& $\langle \hat{g}_{ab}a|\frac{\mathbf{r}\times\mathbf{p}}{r^3}|b\rangle^S$
        & $\langle \hat{g}_{ab}\sigma\mathbf{p}a|\frac{\mathbf{r}\times\sigma}{r^3}|b\rangle^J$       \\
& $(\hat{g}_{ab}a~b|cd)^S$
        & $(\hat{g}_{ab}a~b|cd)^J$                                                                    \\
&       & $(\hat{g}_{ab}\sigma\mathbf{p}a~\sigma\mathbf{p}b|cd)^J$                                    \\
&       & $(\sigma\mathbf{p}a~\sigma\mathbf{p}b|\hat{g}_{cd} c~d)^J$                                  \\
&       & $(\hat{g}_{ab}\sigma\mathbf{p}a~\sigma\mathbf{p}b|\sigma\mathbf{p}c~\sigma\mathbf{p}d)^J$   \\
&       & $(\mathbf{r}\times\sigma a~\sigma\mathbf{p}b|cd)^J$                                                         \\
&       & $(\mathbf{r}\times\sigma a~\sigma\mathbf{p}b|\sigma\mathbf{p}c~\sigma\mathbf{p}d)^J$                        \\
    \hline
  \end{tabular}
  \label{tab:ints}
\end{table}

Table \ref{tab:crco6} and \ref{tab:uf6} are the results of the ground HF
energies analytical nuclear gradients and the isotopic NMR shielding constants
for both the non-relativistic and the 4C Dirac-Coulomb relativistic Hamiltonian.

By fixing the C-O bond at 1.140 \AA\cite{Ziegler1997},
we optimized the geometry, particularly, the Cr-C bond length in terms of the
HF nuclear gradients.
The equilibrium Cr-C bond length based on the non-relativistic Hamiltonian is
2.0106 \AA.
The relativistic effects strengthen the Cr-C bond and shorten it to 1.998 \AA.
On the contrary, the relativistic effects increase the U-F bond length from
1.977 \AA to 1.983 \AA.

The non-relativistic total shielding for chromium is $-5718.5$ ppm, which is far
below the DFT value 507 ppm\cite{Ziegler1997a}.
In contrast to the observation of the ZORA (zeroth order regular approximation)
DFT simulation\cite{Ziegler1997a} which found that the relativistic effects
decrease the shielding constants, our 4C RMB-GIAO computation increases the
shielding to -4622.5 ppm.
Note that the HF exchange is a big source of the paramagentism.
When we switch off the HF exchange in the coupled perturbation Hartree-Fock
solver, the total shielding becomes 999.7 ppm in the 4C relativistic theory
and 937.2 ppm in the non-relativistic theory.
As expected, the shielding constants provided by DFT simulation lie between the
full-exchange and the none-exchange limits.
For uranium in UF$_6$ molecule, we can observe the similar trend that the
relativistic effects increase the para-magnetism.
An interesting phenomenan is the huge increment due to the relativistic
effects which even changes the sign of the total magnetic shielding parameter, from
the deshielding effect in the non-relativistic theory to a shielding effect.

\begin{table}
  \centering
  \caption{Hartree-Fock ground state energy, nuclear gradients (for
  Cr-C bond length, C-O bond was fixed at 1.140\AA), and NMR shielding
  constants of Cr(CO)$_6$. }
  \begin{threeparttable}
  \begin{tabular}{ccccc}
    \hline
                       & Non-relativistic & Dirac-Coulomb & ref \\
    \hline
    \multicolumn{4}{l}{Energy gradients with respect to Cr-C bond length} \\
           1.996 \AA   &                  &$ 0.000464  $  & \\
           1.998 \AA   &                  &$-0.000087  $  & \\
           2.000 \AA   &                  &$-0.000286  $  & \\
           2.008 \AA   &$0.000453  $      &               & \\
           2.010 \AA   &$0.000096  $      &               & \\
           2.012 \AA   &$-0.000258 $      &               & \\
         $r_e$ / \AA   &$2.0106    $      &$ 1.998     $  & 1.918\tnote{a}, 1.998\tnote{b}\\
    HF energy / au     &$-1719.9158$      &$ -1729.8110$  & \\
    NMR shielding / ppm&                  &               & \\
    Cr $\sigma^{dia} $ &$1815.0    $      &$ 1819.2    $  & 1801\tnote{c} \\
    Cr $\sigma^{para}$ &$-7533.5   $      &$ -6441.7   $  & $-2419$\tnote{c} \\
    Cr $\sigma^{tot} $ &$-5718.5   $      &$ -4622.5   $  & \\
    \hline
  \end{tabular}
  \begin{tablenotes}
  \item[a] Experimental data from reference \nocite{CrCbond}\citenum{CrCbond}.
  \item[b] Reference \nocite{Lindh1993a}\citenum{Lindh1993a}.
  \item[c] Reference \nocite{Ziegler2000}\citenum{Ziegler2000}. All electron
    ZORA with TZ/QZ basis.
  \end{tablenotes}
  \end{threeparttable}
  \label{tab:crco6}
\end{table}

\begin{table}
  \centering
  \caption{Hartree-Fock ground state energy, nuclear gradients (for U-F bond
  length) and NMR shielding constants of UF$_6$.}
  \begin{threeparttable}
  \begin{tabular}{ccccc}
    \hline
                       & Non-relativistic & Dirac-Coulomb & ref \\
    \hline
    \multicolumn{4}{l}{Energy gradients with respect to U-F bond length} \\
           1.976 \AA   &$ 0.000851 $      &               & \\
           1.977 \AA   &$ 0.000019 $      &               & \\
           1.978 \AA   &$-0.000335 $      &               & \\
           1.982 \AA   &                  &$ 0.000499 $   & \\
           1.983 \AA   &                  &$-0.000233 $   & \\
           1.984 \AA   &                  &$-0.000961 $   & \\
         $r_e$ / \AA   &$1.977     $      &$ 1.983     $  & 1.996\tnote{a} \\
    HF energy / au     &$-26260.9964$     &$-28665.8851$  & \\
    NMR shielding / ppm&                  &               & \\
    U $\sigma^{dia} $ &$ 11720.6$         &$12291.9$      & \\
    U $\sigma^{para}$ &$-50791.1$         &$706.6  $      & \\
    U $\sigma^{tot} $ &$-39070.5$         &$12998.5$      & \\
    \hline
  \end{tabular}
  \begin{tablenotes}
  \item[a] Reference \nocite{Schreckenbach2005}\citenum{Schreckenbach2005}.
    ZORA with ECP.
  \end{tablenotes}
  \end{threeparttable}
  \label{tab:uf6}
\end{table}

\section{Performance}
\label{performance}
Performance is an essential feature for an integral package.
The code of Libcint was intensively optimized for computational efficiency.
The optimization includes but is not limited to reusing the intermediates,
improving the CPU cache hits, reducing the overhead of function calls,
using the sparsity of the transformation matrices.
Most of the optimization techniques have already been discussed in
Ref \nocite{Flocke2008}\citenum{Flocke2008}.
Besides, we fixed the memory addresses for most intermediates and stored the
addresses in a lookup table.
It significantly reduced the CPU addressing time of the high dimension arrays.
We didn't adopt the early-contraction scheme as HGP method proposed.
Instead, the recurrence relations of DRK's original formula\cite{Rys1983} was
used in the code.
Although more FLOPS are needed in our implementation, this choice has
advantages on the modern computer architecture.
Comparing to the early-contraction scheme,
intermediate components $I_x$, $I_y$ and $I_z$ of DRK's algorithm require
less number of variables.
As such, more data can be loaded in L1 and L2 cache which reduces the memory
access latency.
Besides that, $I_x$, $I_y$ and $I_z$ are more local and aligned in memory.
It enabled us to use SSE instructions to parallelize the compute-intensive
inner product \eqref{eq:rysum}.
We found that SSE3 instructions provide 10-30\% performance improvements
(Table \ref{tab:molpro}).
In this regard, SSE3 is always enabled in the following tests.

The performance was measured on ethane molecule at geometry of
$R_\mathrm{C-C}=1.54$ \AA, $R_\mathrm{C-H}=1.09$ \AA.
The tests were carried out on a machine of Intel Core-i5 @ 3.1 GHz 4-core CPU
with GCC 4.4.5 and Intel MKL library 10.3 installed.
Libcint library was compiled at \verb$-O3 -msse3$ level optimization.
As a reference, Molpro-2012\cite{MOLPRO2012} (using integral package
SEWARD\cite{Lindh1991}) and Psi4\cite{PSI4} (using integral package
Libint\cite{Libint2}) on the same machine were compiled with \verb$gcc -O3$
and linked against Intel MKL library with AVX instruction activated.

\begin{table}
  \centering
  \caption{CPU time (in seconds) of computing ERI for ethane molecule.}
  \begin{tabular}{lrrrrrr}
    \hline
    basis       & Basis size & Psi4 & Molpro & \multicolumn{2}{c}{Libcint} \\
                &            &      &        &w/o SSE3 & w/ SSE3 \\
    \hline
    6-31G       &   30   &  0.10    & 0.09   &  0.09   &  0.07   \\
    6-311G**    &   72   &  0.64    & 0.49   &  0.49   &  0.41   \\
    ANO         &  238   &  2527.6  & 51.13  &  53.59  &  37.78  \\
    cc-pVDZ     &   58   &  0.45    & 0.34   &  0.24   &  0.21   \\
    aug-cc-pVDZ &  100   &  1.87    & 1.18   &  1.02   &  0.85   \\
    cc-pVTZ     &  144   &  4.98    & 4.82   &  2.65   &  2.05   \\
    aug-cc-pVTZ &  230   &  26.03   & 23.12  &  12.40  &  9.27   \\
    cc-pVQZ     &  290   &  81.24   & 65.12  &  31.51  &  22.60  \\
    aug-cc-pVQZ &  436   &  444.23  & 324.29 &  151.04 &  107.24 \\
    \hline
  \end{tabular}
  \label{tab:molpro}
\end{table}

\begin{figure}[htp]
  \begin{center}
    \includegraphics[width=.9\textwidth]{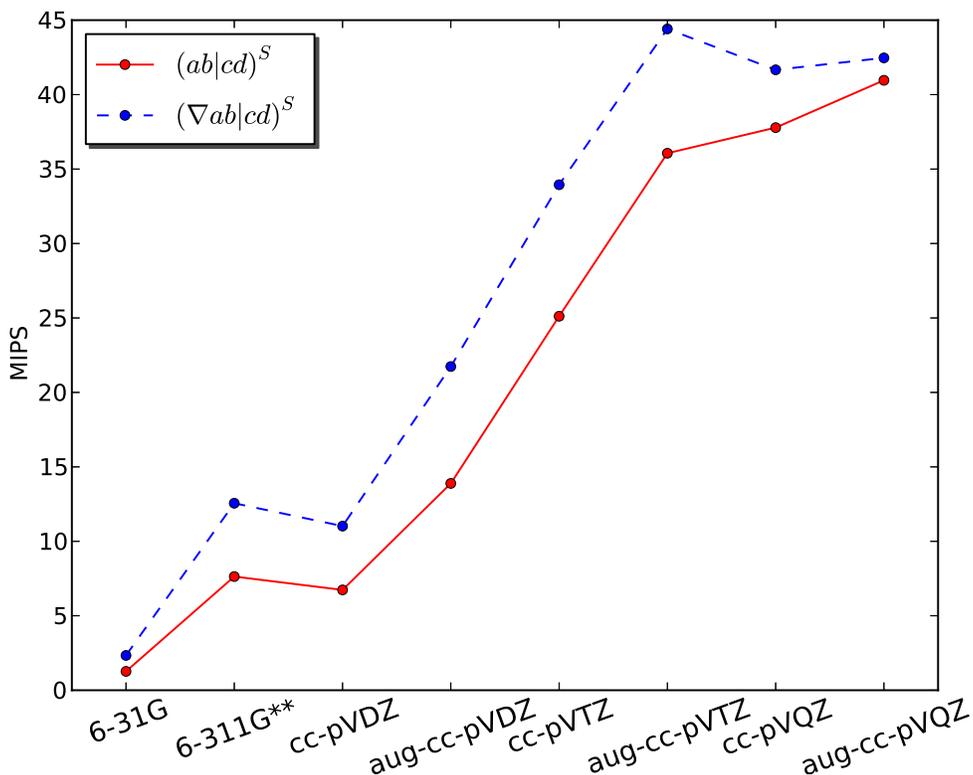}
  \end{center}
  \caption{performance of spherical ERI and ERI gradients.}
  \label{fig:nr-mps}
\end{figure}

Table \ref{tab:molpro} shows the CPU time consumed to compute all basic ERI
integrals $(ab|cd)^S$ (8-fold permutation symmetry was assumed) for double,
triple and quadruple-zeta basis sets (I/O time is not included).
The performance for different basis sets are compared in Figure
\ref{fig:nr-mps}, in which we use MIPS (million integrals per second) to
measure the performance.
In these tests, Libcint library presents high efficiency for the calculation of
the basic ERIs, especially with the loosely contracted basis sets.
In the modern multi-processor computer platform, the overall throughput
can easily exceed the bandwidth that I/O is able to provide.
We tested C$_{60}$ molecule with cc-pVTZ basis (1800 basis functions) on a
cluster of 20 CPU cores running @ 2.5GHz.
It takes 1321 seconds to generate all (1.3 million million) integrals without
using Schwarz inequality, which implies an average bandwidth 7.9 GB/s.
In comparison, the bandwidth of disk or network is typically less than 1
GB/s; GPU to CPU data transfer through PCI-express bus is roughly 10 GB/s.

Integrals other than the basic ERI are implemented by the code generator.
The performance of the spherical ERI gradients $(\nabla a~b|cd)^S$ on a
single CPU core can be found in Figure \ref{fig:nr-mps}.
The performance of integral gradients is better than the basic
spherical ERI in the sense that it owns higher MIPS.
In 6-311G** and cc-pVDZ bases, the gradients are 67 \% (12.5 MIPS vs 7.5
MIPS) and 70 \% (10.9 MIPS vs 6.4 MIPS) faster than the basic ERI.
In the rest cases, the gradients are 10\% - 50 \% faster.

\begin{figure}[htp]
  \begin{center}
    \includegraphics[width=.9\textwidth]{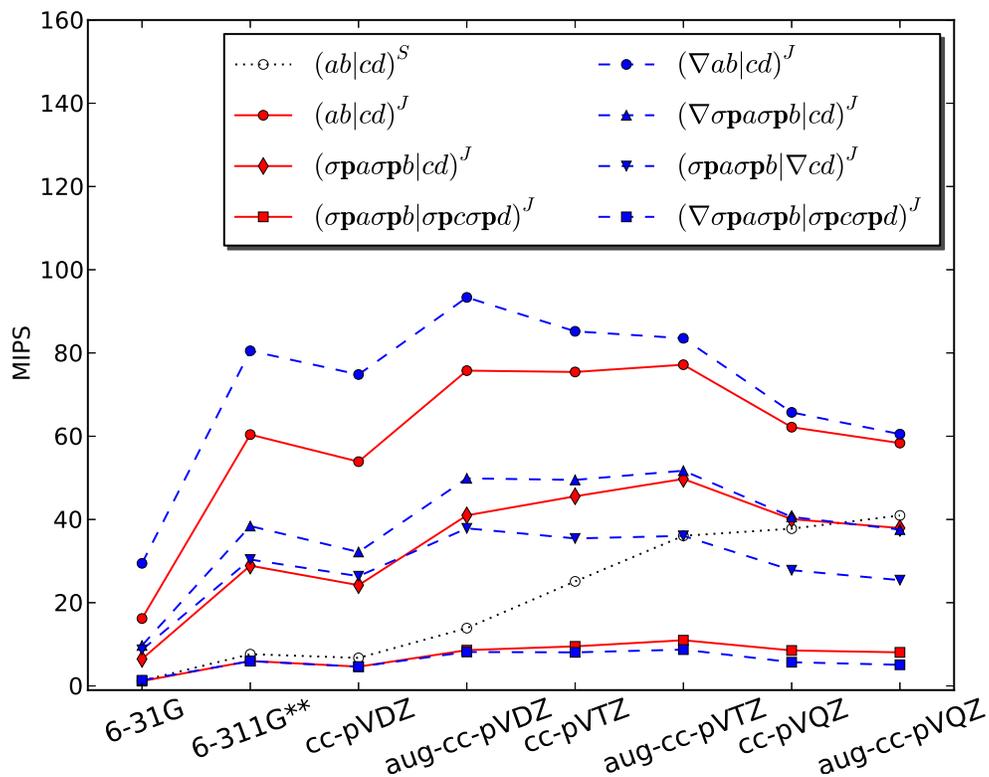}
  \end{center}
  \caption{performance of RKB spinor ERI and RKB ERI gradients.}
  \label{fig:r-mps}
\end{figure}

\begin{table}[htp]
  \centering
  \caption{Timings of the basic ERI and ERI gradients for spherical functions
  and RKB spinor functions.}
  \begin{threeparttable}
  \begin{tabular}{lrrrrrrr}
    \hline
    & \multicolumn{2}{c}{spherical ERI} &
    & \multicolumn{2}{c}{RKB spinor ERI} \\
    & basic (s) & gradients\tnote{a} &
    & basic\tnote{a} & gradients\tnote{b} \\
    \hline
    6-31G       & 0.07   &  6.5 &&  25.0 &  9.5 \\
    6-311G**    & 0.42   &  7.3 &&  30.8 & 11.4 \\
    cc-pVDZ     & 0.21   &  7.3 &&  34.2 & 11.3 \\
    aug-cc-pVDZ & 0.87   &  7.7 &&  39.6 & 12.1 \\
    cc-pVTZ     & 2.09   &  8.9 &&  65.2 & 13.6 \\
    aug-cc-pVTZ & 9.45   &  9.8 &&  83.2 & 14.5 \\
    cc-pVQZ     & 23.19  & 10.9 && 110.8 & 16.5 \\
    aug-cc-pVQZ & 109.91 & 11.6 && 127.0 & 17.3 \\
    \hline        
  \end{tabular}
  \begin{tablenotes}
  \item[a] Data are normalized to the time of basic spherical ERI.
  \item[b] Data are normalized to the time of basic RKB spinor ERI.
  \end{tablenotes}
  \end{threeparttable}
  \label{tab:rkb}
\end{table}

Next, we turn to the performance of spinor integrals.
There are two times as many $j$-adapted spinor functions as spherical
functions for a given basis set.
For the relativistic theory on top of RKB spinor basis,
this results in totally 64 times of the number of spherical integrals to
compute (16 times from $(ab|cd)^J$, 16 times from
$(\sigma\mathbf{p}a~\sigma\mathbf{p}b|\sigma\mathbf{p}c~\sigma\mathbf{p}d)^J$
and 32 times from $(\sigma\mathbf{p}a~\sigma\mathbf{p}b|cd)^J$).
Figure \ref{fig:r-mps} shows the real output of the RKB spinor ERIs on a single
CPU core.
By counting the total number of integrals needed by RKB basis, we can estimate
the relative costs of RKB over the regular spherical ERIs, as shown by Table
\ref{tab:rkb}.
The overall costs of RKB spinor ERI are 25 - 130 times higher than that of
spherical ERI.
Regarding to the fact that the number of integrals in ERI gradients is 12 times
of the number in basic ERIs,
the performance of RKB ERI gradients is worse than the spherical case, since
the cost ratios for RKB are more than 12 in many tests, while they are all
less than 12 for spherical integrals.
We also noticed that the more high angular momentum functions a basis set owns,
the slower RKB integration tends to be.
It reflects the fact that the costs of the RKB ERI are dominated by the
inner product \eqref{eq:rysum} since the four $\mathbf{p}$ operators of
$(\sigma\mathbf{p}a~\sigma\mathbf{p}b|\sigma\mathbf{p}c~\sigma\mathbf{p}d)^J$
introduce $3^4=81$ Cartesian intermediates, which implies 81 times of the
costs of inner products.
This number becomes $3^5=243$ for
$(\nabla\sigma\mathbf{p}a~\sigma\mathbf{p}b|\sigma\mathbf{p}c~\sigma\mathbf{p}d)^J$.
As such, the performance decreasing in ERI gradients is more obvious.

\section{Summary}

The open-source library Libcint provides a new tool to implement integrals for
Gaussian type basis functions.
With this library, efficiency can be obtained for both human labor and machine
costs.
By using the built-in symbolic algebra tool, it is simple to implement various
types of new integrals, including but not limited to
\begin{itemize}
  \item arbitrary order of derivatives,
  \item arbitrary expressions on top of operators $\mathbf{p}$, $\mathbf{r}$
    and $\sigma$,
  \item nuclear attraction, Coulomb and Gaunt interaction,
  \item field-dependent basis functions,
  \item both kinetically and magnetically balanced spinor integrals.
\end{itemize}
As the numerical examples demonstrated in the present work, the analytical
gradients and NMR shielding parameters can be programmed in a simple manner
with the integrals generated by the library. 

The generality of Libcint library is achieved without losing machine efficiency.
On the modern multi-core computers, it is easy to gain an overall throughput
being many times of the I/O bandwidth.
Based on the present Libcint library, future works can be carried out at least
in two aspects.
One is to further optimize and port the code for new computer platforms.
Another is to improve the symbolic algebra tool for more integral frameworks
such as density fitting and multi-particle integrals for explicitly correlated
methods.

\section{Acknowledgments}
The author is grateful to Professor Garnet Chan for his generous support on
this project.  The author thanks Dr. Lan Cheng for insightful discussions and
the comments on the manuscript.

\bibliography{refs}

\providecommand*\mcitethebibliography{\thebibliography}
\csname @ifundefined\endcsname{endmcitethebibliography}
  {\let\endmcitethebibliography\endthebibliography}{}
\begin{mcitethebibliography}{58}
\providecommand*\natexlab[1]{#1}
\providecommand*\mciteSetBstSublistMode[1]{}
\providecommand*\mciteSetBstMaxWidthForm[2]{}
\providecommand*\mciteBstWouldAddEndPuncttrue
  {\def\EndOfBibitem{\unskip.}}
\providecommand*\mciteBstWouldAddEndPunctfalse
  {\let\EndOfBibitem\relax}
\providecommand*\mciteSetBstMidEndSepPunct[3]{}
\providecommand*\mciteSetBstSublistLabelBeginEnd[3]{}
\providecommand*\EndOfBibitem{}
\mciteSetBstSublistMode{f}
\mciteSetBstMaxWidthForm{subitem}{(\alph{mcitesubitemcount})}
\mciteSetBstSublistLabelBeginEnd
  {\mcitemaxwidthsubitemform\space}
  {\relax}
  {\relax}

\bibitem[Boys(1950)]{Boys1950}
Boys,~S.~F. \emph{Proc. Roy. Soc. A} \textbf{1950}, \emph{200}, 542\relax
\mciteBstWouldAddEndPuncttrue
\mciteSetBstMidEndSepPunct{\mcitedefaultmidpunct}
{\mcitedefaultendpunct}{\mcitedefaultseppunct}\relax
\EndOfBibitem
\bibitem[Dupuis et~al.(1976)Dupuis, Rys, and King]{MichelDupuis1976}
Dupuis,~M.; Rys,~J.; King,~H.~F. \emph{J. Chem. Phys.} \textbf{1976},
  \emph{65}, 111\relax
\mciteBstWouldAddEndPuncttrue
\mciteSetBstMidEndSepPunct{\mcitedefaultmidpunct}
{\mcitedefaultendpunct}{\mcitedefaultseppunct}\relax
\EndOfBibitem
\bibitem[Rys et~al.(1983)Rys, Dupuis, and King]{Rys1983}
Rys,~J.; Dupuis,~M.; King,~H.~F. \emph{J. Comput. Chem.} \textbf{1983},
  \emph{4}, 154\relax
\mciteBstWouldAddEndPuncttrue
\mciteSetBstMidEndSepPunct{\mcitedefaultmidpunct}
{\mcitedefaultendpunct}{\mcitedefaultseppunct}\relax
\EndOfBibitem
\bibitem[Pople and Hehre(1978)Pople, and Hehre]{Pople1978}
Pople,~J.~A.; Hehre,~W.~J. \emph{J. Comput. Phys.} \textbf{1978}, \emph{27},
  161 -- 168\relax
\mciteBstWouldAddEndPuncttrue
\mciteSetBstMidEndSepPunct{\mcitedefaultmidpunct}
{\mcitedefaultendpunct}{\mcitedefaultseppunct}\relax
\EndOfBibitem
\bibitem[McMurchie and Davidson(1978)McMurchie, and Davidson]{McMurchie1978}
McMurchie,~L.~E.; Davidson,~E.~R. \emph{J. Comput. Phys.} \textbf{1978},
  \emph{26}, 218 -- 231\relax
\mciteBstWouldAddEndPuncttrue
\mciteSetBstMidEndSepPunct{\mcitedefaultmidpunct}
{\mcitedefaultendpunct}{\mcitedefaultseppunct}\relax
\EndOfBibitem
\bibitem[Obara and Saika(1986)Obara, and Saika]{Obara1986}
Obara,~S.; Saika,~A. \emph{J. Chem. Phys.} \textbf{1986}, \emph{84}, 3963\relax
\mciteBstWouldAddEndPuncttrue
\mciteSetBstMidEndSepPunct{\mcitedefaultmidpunct}
{\mcitedefaultendpunct}{\mcitedefaultseppunct}\relax
\EndOfBibitem
\bibitem[Obara and Saika(1988)Obara, and Saika]{Obara1988}
Obara,~S.; Saika,~A. \emph{J. Chem. Phys.} \textbf{1988}, \emph{89},
  1540--1559\relax
\mciteBstWouldAddEndPuncttrue
\mciteSetBstMidEndSepPunct{\mcitedefaultmidpunct}
{\mcitedefaultendpunct}{\mcitedefaultseppunct}\relax
\EndOfBibitem
\bibitem[Schlegel(1982)]{Schlegel1982}
Schlegel,~H.~B. \emph{J. Chem. Phys.} \textbf{1982}, \emph{77},
  3676--3681\relax
\mciteBstWouldAddEndPuncttrue
\mciteSetBstMidEndSepPunct{\mcitedefaultmidpunct}
{\mcitedefaultendpunct}{\mcitedefaultseppunct}\relax
\EndOfBibitem
\bibitem[Klopper and R{\"o}hse(1992)Klopper, and R{\"o}hse]{Klopper1992}
Klopper,~W.; R{\"o}hse,~R. \emph{Theor. Chem. Acc.} \textbf{1992}, \emph{83},
  441--453\relax
\mciteBstWouldAddEndPuncttrue
\mciteSetBstMidEndSepPunct{\mcitedefaultmidpunct}
{\mcitedefaultendpunct}{\mcitedefaultseppunct}\relax
\EndOfBibitem
\bibitem[Head-Gordon and Pople(1988)Head-Gordon, and Pople]{Pople1988}
Head-Gordon,~M.; Pople,~J.~A. \emph{J. Chem. Phys.} \textbf{1988}, \emph{89},
  5777--5786\relax
\mciteBstWouldAddEndPuncttrue
\mciteSetBstMidEndSepPunct{\mcitedefaultmidpunct}
{\mcitedefaultendpunct}{\mcitedefaultseppunct}\relax
\EndOfBibitem
\bibitem[Gill et~al.(1989)Gill, Head-Gordon, and Pople]{Pople1989}
Gill,~P. M.~W.; Head-Gordon,~M.; Pople,~J.~A. \emph{Int. J. Quant. Chem}
  \textbf{1989}, \emph{36}, 269--280\relax
\mciteBstWouldAddEndPuncttrue
\mciteSetBstMidEndSepPunct{\mcitedefaultmidpunct}
{\mcitedefaultendpunct}{\mcitedefaultseppunct}\relax
\EndOfBibitem
\bibitem[Gill et~al.(1990)Gill, Head-Gordon, and Pople]{Pople1990}
Gill,~P. M.~W.; Head-Gordon,~M.; Pople,~J.~A. \emph{J. Phys. Chem.}
  \textbf{1990}, \emph{94}, 5564--5572\relax
\mciteBstWouldAddEndPuncttrue
\mciteSetBstMidEndSepPunct{\mcitedefaultmidpunct}
{\mcitedefaultendpunct}{\mcitedefaultseppunct}\relax
\EndOfBibitem
\bibitem[Gill and Pople(1991)Gill, and Pople]{Pople1991}
Gill,~P. M.~W.; Pople,~J.~A. \emph{Int. J. Quant. Chem.} \textbf{1991},
  \emph{40}, 753--772\relax
\mciteBstWouldAddEndPuncttrue
\mciteSetBstMidEndSepPunct{\mcitedefaultmidpunct}
{\mcitedefaultendpunct}{\mcitedefaultseppunct}\relax
\EndOfBibitem
\bibitem[Lindh et~al.(1991)Lindh, Ryu, and Liu]{Lindh1991}
Lindh,~R.; Ryu,~U.; Liu,~B. \emph{J. Chem. Phys.} \textbf{1991}, \emph{95},
  5889\relax
\mciteBstWouldAddEndPuncttrue
\mciteSetBstMidEndSepPunct{\mcitedefaultmidpunct}
{\mcitedefaultendpunct}{\mcitedefaultseppunct}\relax
\EndOfBibitem
\bibitem[Lindh(1993)]{Lindh1993}
Lindh,~R. \emph{Theor. Chem. Acc.} \textbf{1993}, \emph{85}, 423--440\relax
\mciteBstWouldAddEndPuncttrue
\mciteSetBstMidEndSepPunct{\mcitedefaultmidpunct}
{\mcitedefaultendpunct}{\mcitedefaultseppunct}\relax
\EndOfBibitem
\bibitem[Dupuis and Marquez(2001)Dupuis, and Marquez]{Dupuis2001}
Dupuis,~M.; Marquez,~A. \emph{J. Chem. Phys.} \textbf{2001}, \emph{114},
  2067--2078\relax
\mciteBstWouldAddEndPuncttrue
\mciteSetBstMidEndSepPunct{\mcitedefaultmidpunct}
{\mcitedefaultendpunct}{\mcitedefaultseppunct}\relax
\EndOfBibitem
\bibitem[Ishida(1991)]{Ishida1991}
Ishida,~K. \emph{J. Chem. Phys.} \textbf{1991}, \emph{95}, 5198\relax
\mciteBstWouldAddEndPuncttrue
\mciteSetBstMidEndSepPunct{\mcitedefaultmidpunct}
{\mcitedefaultendpunct}{\mcitedefaultseppunct}\relax
\EndOfBibitem
\bibitem[Ishida(1996)]{Ishida1996}
Ishida,~K. \emph{Int. J. Quant. Chem.} \textbf{1996}, \emph{59}, 209--218\relax
\mciteBstWouldAddEndPuncttrue
\mciteSetBstMidEndSepPunct{\mcitedefaultmidpunct}
{\mcitedefaultendpunct}{\mcitedefaultseppunct}\relax
\EndOfBibitem
\bibitem[Helgaker et~al.(2012)Helgaker, Coriani, J{\o}rgensen, Kristensen,
  Olsen, and Ruud]{Helgaker2012}
Helgaker,~T.; Coriani,~S.; J{\o}rgensen,~P.; Kristensen,~K.; Olsen,~J.;
  Ruud,~K. \emph{Chem. Rev.} \textbf{2012}, \emph{112}, 543--631\relax
\mciteBstWouldAddEndPuncttrue
\mciteSetBstMidEndSepPunct{\mcitedefaultmidpunct}
{\mcitedefaultendpunct}{\mcitedefaultseppunct}\relax
\EndOfBibitem
\bibitem[Ekstr{\"o}m et~al.(2010)Ekstr{\"o}m, Visscher, Bast, Thorvaldsen, and
  Ruud]{Visscher2010}
Ekstr{\"o}m,~U.; Visscher,~L.; Bast,~R.; Thorvaldsen,~A.~J.; Ruud,~K. \emph{J.
  Chem. Theory Comput.} \textbf{2010}, \emph{6}, 1971--1980\relax
\mciteBstWouldAddEndPuncttrue
\mciteSetBstMidEndSepPunct{\mcitedefaultmidpunct}
{\mcitedefaultendpunct}{\mcitedefaultseppunct}\relax
\EndOfBibitem
\bibitem[{London, F.}(1937)]{London1937}
{London, F.}, \emph{J. Phys. Radium} \textbf{1937}, \emph{8}, 397--409\relax
\mciteBstWouldAddEndPuncttrue
\mciteSetBstMidEndSepPunct{\mcitedefaultmidpunct}
{\mcitedefaultendpunct}{\mcitedefaultseppunct}\relax
\EndOfBibitem
\bibitem[Ditchfield(1972)]{Ditchfield1972}
Ditchfield,~R. \emph{J. Chem. Phys.} \textbf{1972}, \emph{56}, 5688\relax
\mciteBstWouldAddEndPuncttrue
\mciteSetBstMidEndSepPunct{\mcitedefaultmidpunct}
{\mcitedefaultendpunct}{\mcitedefaultseppunct}\relax
\EndOfBibitem
\bibitem[Darling and Schlegel(1994)Darling, and Schlegel]{Schlegel1994}
Darling,~C.~L.; Schlegel,~H.~B. \emph{J. Phys. Chem.} \textbf{1994}, \emph{98},
  5855--5861\relax
\mciteBstWouldAddEndPuncttrue
\mciteSetBstMidEndSepPunct{\mcitedefaultmidpunct}
{\mcitedefaultendpunct}{\mcitedefaultseppunct}\relax
\EndOfBibitem
\bibitem[Pyykko(1988)]{Pyykko1988}
Pyykko,~P. \emph{Chem. Rev.} \textbf{1988}, \emph{88}, 563\relax
\mciteBstWouldAddEndPuncttrue
\mciteSetBstMidEndSepPunct{\mcitedefaultmidpunct}
{\mcitedefaultendpunct}{\mcitedefaultseppunct}\relax
\EndOfBibitem
\bibitem[Stanton and Havriliak(1984)Stanton, and Havriliak]{Stanton1984}
Stanton,~R.~E.; Havriliak,~S. \emph{J. Chem. Phys.} \textbf{1984}, \emph{81},
  1910\relax
\mciteBstWouldAddEndPuncttrue
\mciteSetBstMidEndSepPunct{\mcitedefaultmidpunct}
{\mcitedefaultendpunct}{\mcitedefaultseppunct}\relax
\EndOfBibitem
\bibitem[Cheng et~al.(2014)Cheng, Stopkowicz, and Gauss]{Cheng2014}
Cheng,~L.; Stopkowicz,~S.; Gauss,~J. \emph{Int. J. Quant. Chem.} \textbf{2014},
  \emph{114}, 1108--1127\relax
\mciteBstWouldAddEndPuncttrue
\mciteSetBstMidEndSepPunct{\mcitedefaultmidpunct}
{\mcitedefaultendpunct}{\mcitedefaultseppunct}\relax
\EndOfBibitem
\bibitem[Cheng et~al.(2009)Cheng, Xiao, and Liu]{LanCheng2009a}
Cheng,~L.; Xiao,~Y.; Liu,~W. \emph{J. Chem. Phys.} \textbf{2009}, \emph{131},
  244113\relax
\mciteBstWouldAddEndPuncttrue
\mciteSetBstMidEndSepPunct{\mcitedefaultmidpunct}
{\mcitedefaultendpunct}{\mcitedefaultseppunct}\relax
\EndOfBibitem
\bibitem[Breit(1929)]{Breit1929}
Breit,~G. \emph{Phys. Rev.} \textbf{1929}, \emph{34}, 553--573\relax
\mciteBstWouldAddEndPuncttrue
\mciteSetBstMidEndSepPunct{\mcitedefaultmidpunct}
{\mcitedefaultendpunct}{\mcitedefaultseppunct}\relax
\EndOfBibitem
\bibitem[Kutzelnigg and Liu(2000)Kutzelnigg, and Liu]{Kutzelnigg2000}
Kutzelnigg,~W.; Liu,~W. \emph{J. Chem. Phys.} \textbf{2000}, \emph{112},
  3540--3558\relax
\mciteBstWouldAddEndPuncttrue
\mciteSetBstMidEndSepPunct{\mcitedefaultmidpunct}
{\mcitedefaultendpunct}{\mcitedefaultseppunct}\relax
\EndOfBibitem
\bibitem[Dyall and Jr.(1990)Dyall, and Jr.]{Dyall1990}
Dyall,~K.~G.; Jr.,~K.~F. \emph{Chem. Phys. Lett.} \textbf{1990}, \emph{174}, 25
  -- 32\relax
\mciteBstWouldAddEndPuncttrue
\mciteSetBstMidEndSepPunct{\mcitedefaultmidpunct}
{\mcitedefaultendpunct}{\mcitedefaultseppunct}\relax
\EndOfBibitem
\bibitem[Ishikawa et~al.(1983)Ishikawa, Jr., and Sando]{Ishikawa1983}
Ishikawa,~Y.; Jr.,~R.~B.; Sando,~K. \emph{Chem. Phys. Lett.} \textbf{1983},
  \emph{101}, 111 -- 114\relax
\mciteBstWouldAddEndPuncttrue
\mciteSetBstMidEndSepPunct{\mcitedefaultmidpunct}
{\mcitedefaultendpunct}{\mcitedefaultseppunct}\relax
\EndOfBibitem
\bibitem[Xiao et~al.(2007)Xiao, Peng, and Liu]{YunlongXiao2007}
Xiao,~Y.; Peng,~D.; Liu,~W. \emph{J. Chem. Phys.} \textbf{2007}, \emph{126},
  081101\relax
\mciteBstWouldAddEndPuncttrue
\mciteSetBstMidEndSepPunct{\mcitedefaultmidpunct}
{\mcitedefaultendpunct}{\mcitedefaultseppunct}\relax
\EndOfBibitem
\bibitem[Yanai et~al.(2002)Yanai, Nakajima, Ishikawa, and Hirao]{Yanai2002}
Yanai,~T.; Nakajima,~T.; Ishikawa,~Y.; Hirao,~K. \emph{J. Chem. Phys.}
  \textbf{2002}, \emph{116}, 10122--10128\relax
\mciteBstWouldAddEndPuncttrue
\mciteSetBstMidEndSepPunct{\mcitedefaultmidpunct}
{\mcitedefaultendpunct}{\mcitedefaultseppunct}\relax
\EndOfBibitem
\bibitem[Kelley and Shiozaki(2013)Kelley, and Shiozaki]{Kelley2013}
Kelley,~M.~S.; Shiozaki,~T. \emph{J. Chem. Phys.} \textbf{2013}, \emph{138},
  204113\relax
\mciteBstWouldAddEndPuncttrue
\mciteSetBstMidEndSepPunct{\mcitedefaultmidpunct}
{\mcitedefaultendpunct}{\mcitedefaultseppunct}\relax
\EndOfBibitem
\bibitem[Sun et~al.(2011)Sun, Liu, and Kutzelnigg]{QimingSun2011}
Sun,~Q.; Liu,~W.; Kutzelnigg,~W. \emph{Theor. Chem. Acc.} \textbf{2011},
  \emph{129}, 423\relax
\mciteBstWouldAddEndPuncttrue
\mciteSetBstMidEndSepPunct{\mcitedefaultmidpunct}
{\mcitedefaultendpunct}{\mcitedefaultseppunct}\relax
\EndOfBibitem
\bibitem[Titov et~al.(2013)Titov, Ufimtsev, Luehr, and
  Martinez]{AlexeyTitov2012}
Titov,~A.~V.; Ufimtsev,~I.~S.; Luehr,~N.; Martinez,~T.~J. \emph{J. Chem. Theory
  Comput.} \textbf{2013}, \emph{9}, 213--221\relax
\mciteBstWouldAddEndPuncttrue
\mciteSetBstMidEndSepPunct{\mcitedefaultmidpunct}
{\mcitedefaultendpunct}{\mcitedefaultseppunct}\relax
\EndOfBibitem
\bibitem[Asadchev and Gordon(2012)Asadchev, and Gordon]{Gordon2012a}
Asadchev,~A.; Gordon,~M.~S. \emph{J. Chem. Theory Comput.} \textbf{2012},
  \emph{8}, 4166--4176\relax
\mciteBstWouldAddEndPuncttrue
\mciteSetBstMidEndSepPunct{\mcitedefaultmidpunct}
{\mcitedefaultendpunct}{\mcitedefaultseppunct}\relax
\EndOfBibitem
\bibitem[Blelloch et~al.(1994)Blelloch, Hardwick, Sipelstein, Zagha, and
  Chatterjee]{Blelloch1994}
Blelloch,~G.; Hardwick,~J.; Sipelstein,~J.; Zagha,~M.; Chatterjee,~S. \emph{J.
  Parallel Distributed Comput.} \textbf{1994}, \emph{21}, 4 -- 14\relax
\mciteBstWouldAddEndPuncttrue
\mciteSetBstMidEndSepPunct{\mcitedefaultmidpunct}
{\mcitedefaultendpunct}{\mcitedefaultseppunct}\relax
\EndOfBibitem
\bibitem[Yasuda(2007)]{Yasuda2007}
Yasuda,~K. \emph{J. Comput. Chem.} \textbf{2007}, \emph{29}, 334\relax
\mciteBstWouldAddEndPuncttrue
\mciteSetBstMidEndSepPunct{\mcitedefaultmidpunct}
{\mcitedefaultendpunct}{\mcitedefaultseppunct}\relax
\EndOfBibitem
\bibitem[Asadchev et~al.(2010)Asadchev, Allada, Felder, Bode, Gordon, and
  Windus]{Gordon2010}
Asadchev,~A.; Allada,~V.; Felder,~J.; Bode,~B.~M.; Gordon,~M.~S.; Windus,~T.~L.
  \emph{J. Chem. Theory Comput.} \textbf{2010}, \emph{6}, 696--704\relax
\mciteBstWouldAddEndPuncttrue
\mciteSetBstMidEndSepPunct{\mcitedefaultmidpunct}
{\mcitedefaultendpunct}{\mcitedefaultseppunct}\relax
\EndOfBibitem
\bibitem[Luehr et~al.(2011)Luehr, Ufimtsev, Mart\'{i}nez, and J]{Luehr2011}
Luehr,; Ufimtsev,~N.; Mart\'{i}nez,~I.~S.; J,~T. \emph{J. Chem. Theory Comput.}
  \textbf{2011}, \emph{7}, 949\relax
\mciteBstWouldAddEndPuncttrue
\mciteSetBstMidEndSepPunct{\mcitedefaultmidpunct}
{\mcitedefaultendpunct}{\mcitedefaultseppunct}\relax
\EndOfBibitem
\bibitem[Sun()]{Libcint2014}
Sun,~Q. Libcint library. \url{https://github.com/sunqm/libcint.git}, A simple
  SCF program engined by Libcint library can be found in
  \url{https://github.com/sunqm/pyscf.git}\relax
\mciteBstWouldAddEndPuncttrue
\mciteSetBstMidEndSepPunct{\mcitedefaultmidpunct}
{\mcitedefaultendpunct}{\mcitedefaultseppunct}\relax
\EndOfBibitem
\bibitem[Dyall(2002)]{Dyall2002}
Dyall,~K.~G. \emph{Theor. Chem. Acc.} \textbf{2002}, \emph{108}, 335\relax
\mciteBstWouldAddEndPuncttrue
\mciteSetBstMidEndSepPunct{\mcitedefaultmidpunct}
{\mcitedefaultendpunct}{\mcitedefaultseppunct}\relax
\EndOfBibitem
\bibitem[Shiozaki(2013)]{Shiozaki2013}
Shiozaki,~T. \emph{J. Chem. Theory Comput.} \textbf{2013}, \emph{9},
  4300--4303\relax
\mciteBstWouldAddEndPuncttrue
\mciteSetBstMidEndSepPunct{\mcitedefaultmidpunct}
{\mcitedefaultendpunct}{\mcitedefaultseppunct}\relax
\EndOfBibitem
\bibitem[Wang and Li(2002)Wang, and Li]{Wang2002a}
Wang,~F.; Li,~L. \emph{Journal of Computational Chemistry} \textbf{2002},
  \emph{23}, 920--927\relax
\mciteBstWouldAddEndPuncttrue
\mciteSetBstMidEndSepPunct{\mcitedefaultmidpunct}
{\mcitedefaultendpunct}{\mcitedefaultseppunct}\relax
\EndOfBibitem
\bibitem[Komorovsk\'{o} et~al.(2008)Komorovsk\'{o}, Repisk\'{o}, Malkina,
  Malkin, Ondk, and Kaupp]{Komorovsko2008}
Komorovsk\'{o},~S.; Repisk\'{o},~M.; Malkina,~O.~L.; Malkin,~V.~G.;
  Ondk,~I.~M.; Kaupp,~M. \emph{J. Chem. Phys.} \textbf{2008}, \emph{128},
  104101\relax
\mciteBstWouldAddEndPuncttrue
\mciteSetBstMidEndSepPunct{\mcitedefaultmidpunct}
{\mcitedefaultendpunct}{\mcitedefaultseppunct}\relax
\EndOfBibitem
\bibitem[Xiao et~al.(2012)Xiao, Sun, and Liu]{Xiao2012}
Xiao,~Y.; Sun,~Q.; Liu,~W. \emph{Theor. Chem. Acc.} \textbf{2012}, \emph{131},
  1080, 10.1007/s00214-011-1080-z\relax
\mciteBstWouldAddEndPuncttrue
\mciteSetBstMidEndSepPunct{\mcitedefaultmidpunct}
{\mcitedefaultendpunct}{\mcitedefaultseppunct}\relax
\EndOfBibitem
\bibitem[Ehlers et~al.(1997)Ehlers, Ruiz-Morales, Baerends, and
  Ziegler]{Ziegler1997}
Ehlers,~A.~W.; Ruiz-Morales,~Y.; Baerends,~E.~J.; Ziegler,~T. \emph{Inorg.
  Chem.} \textbf{1997}, \emph{36}, 5031--5036\relax
\mciteBstWouldAddEndPuncttrue
\mciteSetBstMidEndSepPunct{\mcitedefaultmidpunct}
{\mcitedefaultendpunct}{\mcitedefaultseppunct}\relax
\EndOfBibitem
\bibitem[Schreckenbach and Ziegler(1997)Schreckenbach, and
  Ziegler]{Ziegler1997a}
Schreckenbach,~G.; Ziegler,~T. \emph{Int. J. Quant. Chem.} \textbf{1997},
  \emph{61}, 899--918\relax
\mciteBstWouldAddEndPuncttrue
\mciteSetBstMidEndSepPunct{\mcitedefaultmidpunct}
{\mcitedefaultendpunct}{\mcitedefaultseppunct}\relax
\EndOfBibitem
\bibitem[Jost and Rees()Jost, and Rees]{CrCbond}
Jost,~A.; Rees,~B. \emph{Acta. Crystallogr.} \emph{B31}, 2649\relax
\mciteBstWouldAddEndPuncttrue
\mciteSetBstMidEndSepPunct{\mcitedefaultmidpunct}
{\mcitedefaultendpunct}{\mcitedefaultseppunct}\relax
\EndOfBibitem
\bibitem[Barnes et~al.(1993)Barnes, Liu, and Lindh]{Lindh1993a}
Barnes,~L.~A.; Liu,~B.; Lindh,~R. \emph{J. Chem. Phys.} \textbf{1993},
  \emph{98}, 3978--3989\relax
\mciteBstWouldAddEndPuncttrue
\mciteSetBstMidEndSepPunct{\mcitedefaultmidpunct}
{\mcitedefaultendpunct}{\mcitedefaultseppunct}\relax
\EndOfBibitem
\bibitem[Bouten et~al.(2000)Bouten, Baerends, van Lenthe, Visscher,
  Schreckenbach, and Ziegler]{Ziegler2000}
Bouten,~R.; Baerends,~E.~J.; van Lenthe,~E.; Visscher,~L.; Schreckenbach,~G.;
  Ziegler,~T. \emph{J. Phys. Chem. A} \textbf{2000}, \emph{104},
  5600--5611\relax
\mciteBstWouldAddEndPuncttrue
\mciteSetBstMidEndSepPunct{\mcitedefaultmidpunct}
{\mcitedefaultendpunct}{\mcitedefaultseppunct}\relax
\EndOfBibitem
\bibitem[Schreckenbach(2005)]{Schreckenbach2005}
Schreckenbach,~G. \emph{International Journal of Quantum Chemistry}
  \textbf{2005}, \emph{101}, 372--380\relax
\mciteBstWouldAddEndPuncttrue
\mciteSetBstMidEndSepPunct{\mcitedefaultmidpunct}
{\mcitedefaultendpunct}{\mcitedefaultseppunct}\relax
\EndOfBibitem
\bibitem[Flocke and Lotrich(2008)Flocke, and Lotrich]{Flocke2008}
Flocke,~N.; Lotrich,~V. \emph{J. Comput. Chem.} \textbf{2008}, \emph{29},
  2722--2736\relax
\mciteBstWouldAddEndPuncttrue
\mciteSetBstMidEndSepPunct{\mcitedefaultmidpunct}
{\mcitedefaultendpunct}{\mcitedefaultseppunct}\relax
\EndOfBibitem
\bibitem[Werner et~al.(2012)Werner, Knowles, Knizia, Manby, {Sch\"{u}tz},
  Celani, Korona, Lindh, Mitrushenkov, Rauhut, Shamasundar, Adler, Amos,
  Bernhardsson, Berning, Cooper, Deegan, Dobbyn, Eckert, Goll, Hampel,
  Hesselmann, Hetzer, Hrenar, Jansen, K\"oppl, Liu, Lloyd, Mata, May,
  McNicholas, Meyer, Mura, Nicklass, O'Neill, Palmieri, Peng, Pfl\"uger,
  Pitzer, Reiher, Shiozaki, Stoll, Stone, Tarroni, Thorsteinsson, and
  Wang]{MOLPRO2012}
Werner,~H.-J. et~al.  MOLPRO, version 2012.1, a package of ab initio programs.
  2012; see http://www.molpro.net\relax
\mciteBstWouldAddEndPuncttrue
\mciteSetBstMidEndSepPunct{\mcitedefaultmidpunct}
{\mcitedefaultendpunct}{\mcitedefaultseppunct}\relax
\EndOfBibitem
\bibitem[Turney et~al.(2012)Turney, Simmonett, Parrish, Hohenstein,
  Evangelista, Fermann, Mintz, Burns, Wilke, Abrams, Russ, Leininger, Janssen,
  Seidl, Allen, Schaefer, King, Valeev, Sherrill, and Crawford]{PSI4}
Turney,~J.~M. et~al.  \emph{WIREs: Comput. Mol. Sci.} \textbf{2012}, \emph{2},
  556--565\relax
\mciteBstWouldAddEndPuncttrue
\mciteSetBstMidEndSepPunct{\mcitedefaultmidpunct}
{\mcitedefaultendpunct}{\mcitedefaultseppunct}\relax
\EndOfBibitem
\bibitem[Valeev()]{Libint2}
Valeev,~E.~F. Libcint library.
  \url{http://www.chem.vt.edu/chem-dept/valeev/libint/}\relax
\mciteBstWouldAddEndPuncttrue
\mciteSetBstMidEndSepPunct{\mcitedefaultmidpunct}
{\mcitedefaultendpunct}{\mcitedefaultseppunct}\relax
\EndOfBibitem
\end{mcitethebibliography}

\end{document}